# New Technology, New Rules for Journalism and a New World of Engagement

Loup M. Langton
Western Kentucky University, Bowling Green, USA

Mercedes L. de Uriarte
University of Texas, Austin, USA

Kim Grinfeder
University of Miami, Florida, USA

Paulo N. Vicente
Universidade NOVA de Lisboa, Portugal

## Abstract

*The ways in which people learn, communicate and engage in discussion have changed profoundly during the past decade. As Jenkins related in her book, The Convergence Crisis: An Impending Paradigm Shift in Advertising, Millenials do not want to be told the whole story. Rather, they want someone to begin a conversation that will engage others to become participants in the development of that story (2015). Technology now allows that to happen, sometimes with unintended and/or ill consequences, but technology also generates a dynamic potential to create international and interactive discourse aimed at addressing shared global challenges.*

## Introduction

We now live in a period of shifting norms and expectations where those previously taken for granted seem fragile. In part, this perception advances because the speed of change, fueled by emerging technology, affects our most basic priorities -- where we live, how we stay informed and how we govern in times of crisis.

The first decades of the 21$^{st}$ century have disrupted these intersecting dynamics. Neighborhood stability felt the impact of gentrification globally. Reliable production of the day's news came under siege as digital capacity expanded and online options increased. Before long European newspapers, like those in the United States, accepted the need to upgrade equipment and seek tech-savvy reporters and editors. Soon came a struggle to hold on to subscribers. That problem, too, became a global concern.

By 2014 *The Guardian* warned that newspapers were "in free fall." They cautioned, "Print editions are being discontinued. Editors are being replaced. Financial losses are mounting. Digital strategies are yet to bear fruit. New Readerships are fickle, promiscuous and hard to impress…. *El País*, which was selling 400,000 copies in 2007, slumped to 267,000 in April. Advertising income plummeted by 60%" (Penketh, Olterman, & Burgen, 2014, p. 1). In France and Germany the situations were similar. American newspapers grappled with comparable conditions.

Newsroom crisis brought a sobering example of a "trickle-down effect." The disruptions required that journalism education also change. College departments have scrambled to reboot to make way for new technology, to find young professors for whom the new communication systems are most familiar. Curriculum has also moved into the new era. These urgencies have coincided with changes within the Association for Education in Journalism and Mass Communication (AEJMC) where a need to internationalize in response to globalization of these fields was evident. The development of the World Journalism Education Congress (WJEC) opened the way for an "unprecedented meeting of journalism education associations and educator/trainers from around the globe to improve journalism education, and therefore journalism worldwide" (Goodman & Steyn, 2017, Preface). As a first step the new alliances undertook a massive project to document global journalism education. The result was a 470-page report published by the Knight Center for Journalism in the Americas at the University of Texas (Goodman & Steyn, 2017).

Meanwhile, newspapers, long dependable sources of internships and jobs for trained, aspiring journalists, began downsizing as subscribers and casual customers fell away. Creators and users at first believed that digital services would become a secure underpinning of democracy, but free access to unrestricted Internet communication soon introduced negative, even hateful messages, the sources of which could be hidden. More recently, a steady and unrelenting period of civic stress fueled by government accusations of "fake news" has surrounded efforts by professional journalists and journalism organizations to seek new venues for news distribution and new means of news production in answer to the digital transformation.

These claims of fake news emerged against familiar sources and circles previously taken for granted as authorities, but led by U. S. President Donald J. Trump, these flourishing attacks have generated insecurity about content. That has encouraged the popularity of social media despite the fact that its users often have no way to discern its content origin or validity. The strategy to undermine quickly became an international approach as world leaders have followed Trump's example from microphone to tweet. This new appropriation of information has made even local reporting difficult: Who to trust as sources? As the need to be well informed about personal, economic and civic issues have increased, traditional, legacy and research sources have fallen victim to recurring attacks. Given the symbiotic history between journalism education and the job-market newsroom, the steady attacks against the press have affected journalism classrooms.

This has caused both opportunity and consternation for journalism educators. Most students now begin studying journalism with a grasp of digital potentials, but reliable sourcing and background relevance as well as an appreciation for context often challenge emerging journalists. Professors may understandably wonder, "how and what then do we teach?" "Where do we find sources more apt to provide valuable and knowledgeable background context to proposed news stories?" "How do we make that information appealing to the new information consumers living in high-tech equipped homes?" "How do we make it pay?" Too long hamstrung by mythologies of journalistic objectivity or neutrality, reporters have been expected to cultivate a certain professional distance from issues that they may know more about than they report. These may be goals worthy of reconsideration.

Certainly when covering communities that face disappearing affordable housing, reporters should have an intimate knowledge of the area or be embedded there to cover the critical issues of the changes. Unless one has seen presence, it is impossible to see the impact of absence. One approach to reimagining journalism education might have students volunteer at a grassroots-level with a non-profit of their choice for a more ethnographic experience (deUriarte, 1993). This approach might be thought of as revisiting the "beat" concept where reporters and sources interact over time in matters as personal and life determining as one's home.

Another pedagogical approach might be to pursue, as with the project that is the focus of this paper's research, an interactive-technology design that allows various informed perspectives on the issues of affordable housing and gentrification to receive equal voice and to be shared internationally thus fostering worldwide and diverse conversations on the specified topics.

Still, competing with the attractiveness of options that give visual movement to the affordable housing and gentrification story is the need to encourage students to develop interest, sensitivity and empathy to problems faced by the growing number of those disenfranchised. People without place, feel a loss of identity and purpose that may seem remote from student reality. Students know that they are temporary migrants, present only until they claim their pending degrees, but youth usually welcomes new destinies as dreams are pursued. Not so for those cozy in older predictable realities. Many such threatened residents wish to remain where they are.

Whether employing a traditional journalistic approach or one that is tech-driven, these situations provide opportunity for insightful narrative reporting. This requires developing a lens that crosses class and generations, a challenge even for many experienced reporters and editors.

Currently the issue of secure shelter, at-risk reliable reporting and journalism education appear entwined in ways that necessitate exploratory analysis and some consideration of new practices (Guajardo, Guajardo, Janson & Militello, 2016).

**Journalism' Response to Social Change from the Middle of the 20th Century Onward**

Historically, significant social change has fostered analysis of press relevance, compliance and weaknesses. Each one builds upon records of earlier results. The modern journalism era began about 75 years ago as the end of WWII approached. At that time U.S. journalism had gained significant international stature. Coverage of Europe fell to the U.S. wire services because the war limited previous press leaders, Agence France Presse and the BBC's ability to cover and deliver the news. The U.S. press, restricted by censorship in the interest of national security, found itself questioning whether press freedom could be restored. To assure its return and improve its future Henry R. Luce (creator of *Time Magazine*) and Robert Hutchins (president of the University of Chicago) joined forces to create a committee to explore the matter and develop guidelines to assure a professional press.

The resulting Commission on Freedom of the Press (the Hutchins Commission) examined the civilian role of journalism and set five required standards of professional ethics. The Commission presented the key elements as truthful, comprehensive, intelligent, contextualized content and a "representative picture of constituent groups in the society" (Commission on Freedom of the Press, 1947, p. 26). For the first time minorities and foreigners were cited as entitled to inclusive representative coverage. "When the images they (journalists) portray fail to present the social group truly, they tend to pervert judgment" (Commission on Freedom of the Press, 1947, p. 26).

They began the press project in 1944 and published the conclusions in the 1947 committee report *A Fair and Responsible Press*. The Report established five ethical standards to which the Commission believed the press should adhere and that content should provide:

- A truthful, comprehensive and intelligent account of the day's events in a context that gives them meaning

- A forum for the exchange of comment and criticism

- The projection of a representative picture of constituent groups in the society

- The presentation and clarification of the goals and values of the society

- Full access of the days intelligence

The Commission strongly supported a free press as "essential to political liberty." "Where men cannot freely convey their thoughts to one another, no freedom is secure…. Free expression is therefore unique among liberties: it promotes and protects all the rest…. Civilized society is a working system of ideas. It lives and changes by the consumption of ideas" (Commission on Freedom of the Press, 1947, p. 6).

Twenty years later President Lynden Johnson created the National Advisory Commission on Civil Disorders to determine the cause of the more than one hundred uprisings across the U.S. The resultant Kerner Report (released in 1968) did not mince words. They faulted the press for its inaccurate, exaggerated reporting, and they noted that press coverage needed to be representative of all citizens (The National Advisory Commission on Civil Disorders, 2016).

The most influential path toward communication globalization may be attributed to the UNESCO report, *Many Voices One World* (UNESCO, 1980). The report, prepared by the "International Commission for the Study of Communication Problems," chaired by Sean MacBride, criticized media concentration, commercialization and unequal access to information in a sweeping analysis of the "developed world's" media advantage and power to maintain exclusion. It urged universal freedom of expression not vested in just journalists and governments. Instead, it sought a journalistic experience that it defined as "horizontal," one that would replace the existing structure that it believed to be "little more than a dominating monolog."

The MacBride Report further criticized the information flow and content of news noting that the distribution was intricately entwined with the international power structure. The complaints included, the promotion of "alien attitudes across cultural frontiers" (UNESCO, 1980, p. 87) and marketing practices that were "imposing uniformity of taste, style and content" (UNESCO, 1980, p. 88). Moreover, it pointed to the control of communication equipment as a primary means of excluding participation in world dissemination of other points of view and circumstances.

The United States, however, saw the MacBride Report as antagonistic to "the free flow of information" and, along with the United Kingdom and Singapore, withdrew its support from UNESCO in 1984 (it rejoined in 2003). The United States' withdrawal and the subsequent financial consequences put pressure upon UNESCO to distance itself from the MacBride Report, and many of the Report's recommendations fell away.

Most recently the 467-page publication, *Global Journalism Education In the 21st Century: Challenges &* Innovations, gathered the work of journalism education scholars discussing the titled topic as it pertains to preparation for a journalism profession as practiced in ten selected nations. The book includes a significant contribution by Mark Deuze in an "analysis of how globalization and new technologies/innovations are dramatically altering journalism and, accordingly, journalism educators' roles" (Goodman, & Steyn, 2017, p. ix).

**Technology and Change**

This paper's opening statement read, "We now live in a period of shifting norms and expectations where those previously taken for granted seem fragile. In part, this perception advances because the speed of change, fueled by emerging technology, affects our most basic priorities -- where we live, how we stay informed and how we govern in times of crisis." It seems a bit foreboding and inevitable. Yet, the paper has also put forth that "historically, significant social change has fostered analysis of press relevance, compliance and weaknesses. Each one builds upon records of earlier results." Although emerging, interactive technology has powered global disruptions in the ways in which information is exchanged, used and sometimes abused, there is also a recognition by many that the interactive and global characteristics of today's media can be employed in the service of finding solutions (Halliday, 2013; IPECC, n.d.; PHYS, 2015; Wingfield, 2015; WHO, 2014).

Ash Carter began his 2018 Ernst May Lecture to the Aspen Strategy Group with the following observation:

> Disruptive scientific and technological progress is not to me inherently good or inherently evil. But its arc is for us to shape. Technology's progress is furthermore in my judgment unstoppable. But it is quite incorrect that it unfolds inexorably according to its own internal logic and the laws of nature. My experience and observation is that this is true only directionally. Which specific technologies develop most quickly is heavily shaped by the mission that motivates and rewards the innovators: improving health, selling advertising or some other service, cheap energy, education, or national defense, for example.

Carter concludes, however, with a note of caution:

> There are a lot of smart kids at MIT and around Boston working on the driverless car. LIDAR (light detection and ranging), which along with passive imagery and radar provides inputs to the steering algorithms…. I always say to these smart kids, "save a little bit of your innovative energy for the following challenge: How about the carless driver? What is to become of the tens of thousands of truck, taxi, and car drivers whose jobs are disrupted" (2018)?

Certainly the rapidity with which issues such as "gentrification" and "lack of affordable housing" have spread, can be linked to the upsurge in global, social media industries such as Airbnb (Shah, 2017), and so we return to an assertion found earlier in this paper: "Currently the issue of secure shelter, at-risk reliable reporting and journalism education appear entwined in ways that necessitate exploratory analysis and some consideration of new practices." The Lisbon Beta Project is an effort

495

to provide a model for exploring new definitions for journalism; new approaches to journalism education, and new platforms for addressing universal issues.

**Method**

The Project was produced in four stages: research, production, post-production and interactive design. The first three closely followed a traditional journalism approach to digital storytelling. As a participant at the 2016 Habitat III Conference in Quito, this paper's lead author became more interested in and more knowledgeable about global urban issues and agendas while separately investigating ways in which technology has fundamentally changed the ways in which people (mostly younger) communicate, tell stories and exchange ideas. The connection of these two "dots" seemed like a natural union, and the idea of exploring the link turned into a successful sabbatical proposal. Research revealed the "crisis" that Lisbon's historical center was experiencing as a failed economy had created a target for wealthy international investors who looked to capitalize on precious, low-cost real estate. Residents, community groups, social activists and eventually the municipal government identified "affordable housing" as Lisbon's most outstanding social issue in the fall of 2016.

This paper's lead author moved to Lisbon in September, 2016, read local, national and international stories about the affordable housing problem in Lisbon and took note of several people identified in the stories, who held various perspectives on the issues. The author contacted eleven of those individuals and then several others who were identified by the original group. In all, 18 people from all sides of the "Affordable Housing" and related "Gentrification" debates in Lisbon were contacted, and eventually 14 agreed to be interviewed on camera. The interviewees included government officials; people displaced by gentrification; community activist leaders; a journalist; a university scholar doing related research, and others.

From the start the project was envisioned as a collaborative endeavor. Research revealed that the recently created iNOVA Media Lab at the Universidade NOVA de Lisboa had already participated in a video collaboration with a documentary production company from the Netherlands. The Media Lab's Founder and Coordinator, Paulo Vicente, agreed to the proposed collaboration, and all of the 14 interviews were conducted in the University's studio and/or on location. A staff member served as cameraman, and a PhD student served as the interview manager and translator. All 14 participants that had agreed to be interviewed took part. The interviews generally lasted from a half hour to an hour. The interviews were conducted over a four-week period.

Several months of post-production editing was undertaken by the lead author upon returning to the U.S.. An undergraduate student from Western Kentucky University completed the sound and color enhancement portion of the post-production work.

The edited interviews were then sent to the third member of the collaborative team, Kim Grinfeder, the Founder and Director for the Interactive Media Program at the University of Miami (Florida). It was this fourth and final stage of the Project that introduced the interactive element to the Lisbon Beta Project.

**Interactive Storytelling Engages Students in Urban Issues Dialog**

Students often feel disconnected from the traditional media methods taught to them at journalism schools today. In fact, most people under 30 do not subscribe to newspapers or habitually tune in to TV and/or radio newscasts (Shearer, 2018). Their day-to-day media landscape offers a personalized, filtered, searchable, and summarized news feed that covers a breadth of subjects albeit with little depth. In addition, when we ceded editorial control to algorithms that optimize for clicks rather than a world view; it has created an echo chamber making it difficult for us to see opposing arguments on a subject. Today's student media consumption habits are vastly different than those of their educators, and while the fundamentals remain the same, how do we reconcile these differences?

The open-source software produced for this project (Doc-Gen) aims to provide a platform that mimics a student's current media environment to address complex, multi-faceted stories such as affordable housing and promote conceptual understanding. The Lisbon Beta Project software, for example, permits viewers to generate their own 2-4 minute documentaries about affordable housing allowing them to filter topics and reconstruct the story at their own pace. The danger of doing this is that students might only receive a narrow view of the topic by only selecting subjects that interest them, but in user tests, students often came back and continuously "reconfigured" their documentary multiple times to view different perspectives or understand different topics.

The current filters offered are "affordable housing," "social conditions," "rentals," "government," "families," "gentrification," "developers," "tourism," "transportation," and "universities." Viewers can select one or combine multiple filters to generate their documentaries. The generated documentaries come from a "clip bank" that was derived from 14 interviews conducted with local stakeholders from Lisbon. The interviewees are from a diverse socio-economic background and from multiple sides of the issues.

The interviews followed a fixed questioning line (all interviewees were asked the same questions) on diverse topics that allowed us to edit the interviews and create the clip bank. The clips were created using quotes that most directly answered the questions and were then tagged with metadata. The clips ranged from 18 seconds to 1:14 in duration. The metadata included applicable keywords. Once the clip bank was formed, the software allowed viewers to generate the documentaries based on their selected topics. The software randomly selected the interviewees that spoke about the keywords chosen.

**Results**

Doc-Gen allowed the site to include 7 hours of interviews in a format that students are accustomed to viewing. Initial user tests on a group of 16 students showed that the students quickly grasped the concept and kept "reconfiguring" the story. The topic of affordable housing is vast and has multiple angles and even a traditional feature-length documentary would only allow viewers to see a small curated view on the topic.

One of our goals with Doc-Gen was to mimic the functionality of news-feeds, from software such as Snapchat and Instagram, that students find recognizable, and at the same time restrict the content to a specific topic from a curated list of subject experts. By operating on a familiar platform (http://lisbon.formativejournalism.org/about), students seemed more inclined to explore a topic rather than passively absorb it.

Another goal of working with Doc-Gen was to create a platform that provoked discussion. Since every student brought his/her own set of experiences to the selection and viewing of the documentaries, discussions about affordable housing at the end of the user test were lively and came from different perspectives. This, in turn, allowed for greater breadth and insight regarding the topics covered as everyone seemed to have received some part of the story that they were particularly interested in since they had self-selected the topics.

While the content used in Doc-Gen was produced by a team of media professionals, with some student assistance, the content could have had more student participation, and the results would have increased engagement and promoted a deeper learning experience. Today's youth not only consume media but are very adept at producing their own content (Ito, Baumer, Boyd, Cody, Herr,…Tripp, 2009). Including a participatory component in future project creation will allow for greater ownership of the topic In addition, as educators and professionals, we might learn a thing or two on how to produce content for automated video feeds.

## Discussion

### *Teaching journalism in the face of a future shock*

In the face of the profound transformations related to the access and use of digital technologies, as well as connected with a restructuring media industry, communication sciences and journalism education in particular appear to be on the verge of a future shock, one that "occurs when you are confronted by the fact that the world you were educated to believe in doesn't exist" (Postman & Weingartner, 1969, p. 14).

There are signs suggesting that students already consider journalism education inadequate when it comes to keeping up with the emerging communication technologies and that the theory-practice antinomy is remarkably limited to comprehensively address the complexity of contemporary digital technologies and their emerging mediation regimes (Ercan, 2017). On the other hand, even if for pragmatic reasons, news media companies seek candidates with solid multimedia skills (Wenger, Owens, & Cain, 2018), established digital journalists – reporters, editors, producers – consider that new entrants into the newsrooms, although well trained in technological skills, are less prepared when it comes to traditional journalistic skills, such as interviewing, critical thinking, and understanding newsworthiness (Ferrucci, 2017).

It becomes thus apparent that contemporary journalism demands a blend of intellectual education and professional training, often with very different quality assessments, as well as of academic research and teaching (e.g. Bromley, Tumber, & Zelizer, 2001; Meier & Schützeneder, 2019; O'Donnell, 2001; Williams, Guglietti, & Haney, 2017). The Lisbon Beta Project is an effort to implement a knowledge transfer hub aimed at bridging these epistemic gaps and occupational boundaries. As such, it can be adequately understood as being based on a problem-based learning/teaching methodology, through which the development and application of intellectual and technical skills are nurtured, allowing for journalism students' identity to be built in the making.

Interdisciplinary collaboration has a key value in the development of future journalism. Although, contemporary communication sciences and journalism education are still very much embedded with theoretical traditions supportive of an occupational boundary work: the social and the cultural world as a strict sociological and anthropological object of study and the material world as the domain of scientists and engineers. We have sustained elsewhere that the multilayered nature and phenomena of digital media can only be properly addressed by a higher level of integration among fields and disciplines, working together to develop new theories, concepts, methods and applications around common problems. To this process towards transdisciplinarity we have called a digital Renaissance (Vicente, 2018). Nevertheless, to operationalize a link between biological, technological, social and cultural aspects of communication is still an exception and very often a matter of academic occupational boundary and contest (Bondebjerg, 2017).

Since emerging digital technologies rearrange the set of skills, operations and mindsets that journalists and journalism students and professors were used to, the integration of several disciplines in a project, as well as in a university course, was a key concern for the Lisbon Beta Project. Innovative dynamics in teaching/learning require a solid epistemic foundation, particularly because Practice-based research (PbR) and Project-based Learning (PbL) in Digital Journalism are still underrepresented in communication university departments.

The epistemic divide between theory and practice in conventional journalism education disregards that practice-based research (PbR) and Project-based Learning (PbL) are founded on a co-evolutionary premise "where the existing technology is used in a new way and from which technology research derives new answers: in turn, the use of new digital technology may lead to transformation of existing forms and traditional practices" (Edmonds et al., 2005, p. 458). To transform our classrooms into laboratories – or into a simulated version of a newsroom – we do need to recognize that while practice generates the relevant artifacts and phenomena, scientific research grants its systematic investigation. Thus, "not only is practice embedded in the research process, but research questions arise from the process of practice, the answers to which are directed towards enlightening and enhancing practice" (Candy & Edmonds, 2018, p. 63).

Active professor and student engagement are critical. As such, a robust framework needs to support student-centered participation. It is relevant to make clear that, while interdependent, practice does not equals research. As a scientific method, PbR demands that researchers "develop theoretical frameworks that inform and guide the making and evaluation of the outcomes of their practice" (Edmonds & Candy, 2010, p. 470), i.e. for the research process it is central and unavoidable both the theoretical and conceptual constructs and working from within digital media to test and advance ideas. In this sense, The Lisbon Beta Project not only aspires to be an artifact, but also a framework to explore a theorist-practitioner model in journalism teaching.

We must realize that the media landscape, where our students consume content, has dramatically changed. Teaching students broadcasting in a traditional sense is the equivalent of teaching them Latin; it's useful to better understand the roots of a particular group of languages but outdated for everyday use. We must adapt to the shifting landscape and look for new ways to give students the tools to explore a subject in depth and from multiple perspectives that match their current media landscape. Media consumption has never been higher, according to the Pew Research Center, 95% of teens have access to a smartphone and 45% say they are on it "near constantly" (Teens, 2018). And, while the fundamentals of journalism remain as critical as ever, developing new tools that can explore complex stories in a manner that is engaging and familiar with this new generation is critical for the survival of journalism education.

**About the Authors**

***Loup Langton*** serves as the Turner Professor at the Western Kentucky University's School of Journalism & Broadcasting. His career reflects a balance between professional and creative work and research with a particular passion for Latin America. His book, Photojournalism and Today's News, continues to receive excellent reviews across the Americas and Europe. Langton and Ecuadorian Pablo Corral created and direct the POY Latam visual journalism contest that is the largest and most prestigious contest of its kind in Ibero America. He received his Doctor of Philosophy degree from the University of Texas School of Communication in 1994.

***Dr. Mercedes de Uriarte*** has long combined presswork and an academic career. She joined the Journalism Department in 1987 and became an affiliate of the Lozano Long Institute for Latin American Studies. In 1991 she became the second woman and first minority ever to be awarded tenure in the then Department of Journalism. Besides numerous articles and book chapters, she is the principal investigator or director of several grant-funded projects.
She is a 2010 recipient of a Kellogg Foundation three-year grant to collect the voices of long-time residents of East Austin displaced by ongoing intense gentrification of the area.

***Kim Grinfeder*** is the Founder and Director for the Interactive Media Program at the University of Miami. Kim likes to take complex ideas and distill them into clear messaging and visual concepts. His production skill set wavers between design and code with a strong focus on user experience. As an entrepreneur Kim has successfully established and sold two companies. He joined the faculty at the School of Communication in 2003 and holds a Master's degree in Interactive Telecommunications from New York University and a Bachelor of Arts in History from the University of Miami.

***Paulo Nuno Vicente*** is an Assistant Professor of digital media at Universidade NOVA de Lisboa (Portugal). He founded and coordinates iNOVA Media Lab, a digital creation laboratory developing research at the convergence of creative digital media and emerging technologies. The lab is anchored around six key thematic areas: immersive and interactive narrative, information visualization, digital methods and web platforms, science communication, digital journalism and the future of education. He was a 2016 Fellow of the German Marshall Fund of the United States. He completed his PhD in Digital Media in 2013, in the scope of UT Austin Portugal Program.